\documentclass[a4paper]{article}

\usepackage{INTERSPEECH2020}
\usepackage{hyperref}

\title{ICE-Talk: an Interface for a Controllable Expressive Talking Machine}

\name{No\'e Tits~\href{https://orcid.org/0000-0002-1971-4412}{\includegraphics[scale=0.7]{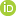}}, Kevin El Haddad~\href{https://orcid.org/0000-0003-1465-6273}{\includegraphics[scale=0.7]{orcid.png}}, Thierry Dutoit}
\address{
  Numediart Institute, University of Mons
  }
\email{\{noe.tits, kevin.elhaddad, thierry.dutoit\}@umons.ac.be }

\begin{document}

\maketitle
\begin{abstract}
    ICE-Talk is an open source\footnote{https://github.com/noetits/ICE-Talk} web-based GUI that allows the use of a TTS system with controllable parameters via a text field and a clickable 2D plot. It enables the study of latent spaces for controllable TTS. Moreover it is implemented as a module that can be used as part of a Human-Agent interaction.
\end{abstract}

\section{Introduction and Motivations}

Speech Synthesis is an important component of Human-Robot Interaction~\cite{theory_behind_control_tts-19-tits}. However as of today, expressiveness in speech generated by Text-to-Speech (TTS) systems is under-explored in such interactions.
The reason is the difficulty of accessing the variables controlling speech expressiveness in a deep learning-based TTS system~\cite{methodology_doctoral_consortium-19-tits}.

To tackle this problem we propose a tool allowing the control of these variables through a graphical interface, thus contributing to the democratization of the use of Deep Learning (DL)-based TTS systems in Human-Agent interaction (HAI) applications.
The goal of this tool is to be generic enough to be connected in any HAI application.

This interface allows for the control over the synthesis parameters of a DL-based model through its latent space directly and intuitively in a graphical way. It therefore allows the implementation of several interesting applications and experiments such as listening tests for the evaluation of such systems thanks to easy prototyping of experiments. An example of experiment is available.
This tool also enables the possibility of studying the impact of expressive synthesized speech in Human-Robot interaction.


\section{Related Work}

As of today, there are some web interfaces allowing the use of DL TTS models\footnote{https://github.com/keithito/tacotron}. They allow to write text, that is sent to the model and get the synthesized speech as an audio object that one can listen. The text is therefore the only control variable that we can access. 

Recently, an interface\footnote{https://github.com/CorentinJ/Real-Time-Voice-Cloning} allowing to give an audio for TTS with speaker characteristics was developed based on the research of Tacotron team~\cite{tacotron3-18-ye, multispeaker_voice_cloning-19-jemine}.
It allows to select a reference audio file and synthesize speech from text imitating the voice of the reference.
It is however not possible to interact with a latent space representing acoustic variability.

In this paper we provide a web interface capable of visualizing and exploring a space of voice expressiveness and synthesize corresponding expressive speech.

\section{Description of ICE-Talk}

\subsection{System architecture}

\begin{figure}[h]
  \centering
  \includegraphics[width=0.8\linewidth]{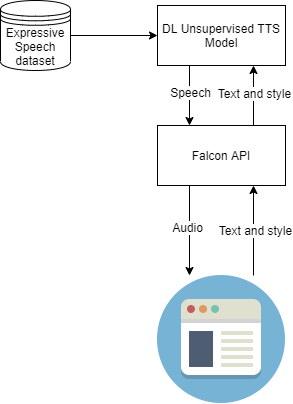}
  \caption{System architecture}
\label{system_architecure}
\end{figure}

Figure~\ref{system_architecure} depicts the different components of the system architecture. It is constituted of a DL unsupervised TTS model trained on an expressive dataset (see Section~\ref{model}).


To make the model available as a web service and communicate information of text, audio and style between the web interface and the TTS model, the Falcon Web framework\footnote{https://falcon.readthedocs.io/en/stable/} is used.

Falcon allows to bridge the gap between a python code and a web interface, allowing the use of Deep Learning frameworks through a web application.

\subsection{DL Unsupervised Model}
\label{model}

As a use case, we use a modified version of \textit{Deep Convolutional Text-to-Speech} (DCTTS)~\cite{dctts-17-tachibana}, a state-of-the art Deep-Learning Sequence-to-Sequence (seq2seq) model with an output controllable through a Latent Space designed to represent variations in voice style as described in~\cite{visualization-19-tits}.

A TTS seq2seq model typically consists of an encoder-decoder structure. Text is encoded as a latent representation that is then decoded with an attention based decoder to predict a mel-spectrogram later inverted to an audio waveform.

To obtain a voice style representation in~\cite{visualization-19-tits}, a mel-spectrogram encoder is added. It consists of a stack of 1D convolutional layers, followed by an average pooling. This operation ensures to obtain time-invariant information. It can thus contain information about statistics of prosody such as pitch average, average speaking rate, but not a pitch evolution.

\subsection{Web Interface}

The interface contains a 2D representation of a latent space which is an internal representation of the data distribution by the network. This 2D representation is obtained via a dimensionality reduction applied to the highly dimensional latent space of the system. The interface also contains a text box for the system's input and an audio player for the system's output.

The latent space represents the distribution of some controlling parameters (the expressiveness for instance) of the output speech, and is obtained after training.

By writing a text and clicking on a point on the 2D space, an audio signal is generated with the parameters values corresponding to the point clicked on.

The web interface is implemented in HTML5 and javascript to use the service.

There are several possibilities for dimensional reduction : UMAP, PCA or t-SNE.
The click of the mouse is detected using javascript in pixels coordinates and mapped to the reduced data space.

Then Nearest Neighbour regression is used to compute the 2D data point, and a lookup table gives the corresponding 8D point of the latent space.
The text and the 8D vector are fed to the model that generates the sentence and save it into a wav file.
The audio wav file is then served and played as an HTML5 audio object.

\begin{figure}[h]
  \centering
  \includegraphics[width=1\linewidth]{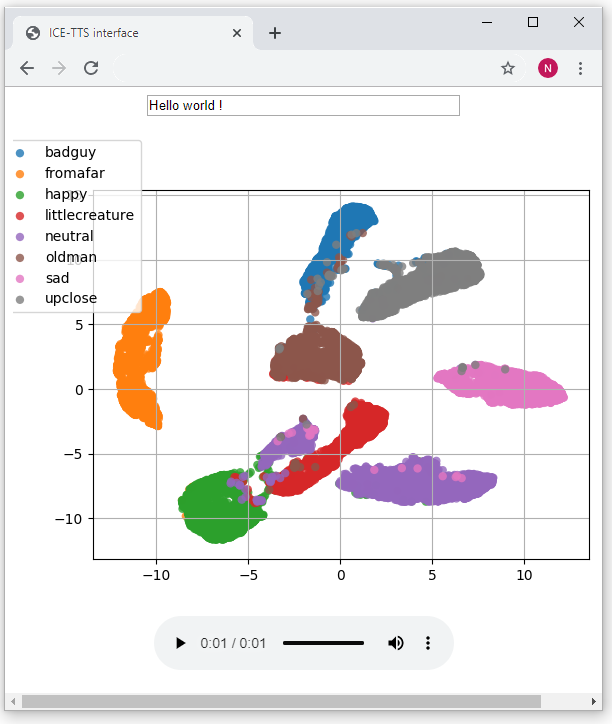}
  \caption{ICE-Talk web interface. It is constituted of a text field, an image representing a latent space of vocal variability contained in the dataset, and an audio player to listen to the synthesized utterance.}
\end{figure}

 

\section{Conclusions and Future Works}

We presented an innovative interface for Controllable Expressive TTS called ICE-Talk that show a proof of concept of research results previously presented at Interspeech. It is open source and ready to be used with available pre-trained models.

As future works, other models available such as a multi-speaker TTS\footnote{https://github.com/CorentinJ/Real-Time-Voice-Cloning} could be integrated to be able to generate speech from different speakers, not based on references but on the latent space describing speaker characteristics.

\section{Acknowledgements}

No\'e Tits  is funded through a FRIA grant (Fonds pour la Formation \`a la Recherche dans l'Industrie et l'Agriculture, Belgium).

\bibliographystyle{IEEEtran}

\bibliography{biblio}

\end{document}